\documentclass[%
reprint,
superscriptaddress,
amsmath,amssymb,
aps,
]{revtex4-1}

\usepackage{color}
\usepackage{graphicx}
\usepackage{dcolumn}
\usepackage{bm}
\usepackage{textcomp}

\begin {document}

\title{Estimating relaxation times from a single trajectory}

\author{Takuma Akimoto}
\email{takuma@rs.tus.ac.jp}
\affiliation{%
  Department of Physics, Tokyo University of Science, Noda, Chiba 278-8510, Japan
}%

\author{Eiji Yamamoto}
\affiliation{%
  Department of System Design Engineering, Keio University, Yokohama, Kanagawa 223-8522, Japan
}%

\author{Takashi Uneyama}
\affiliation{%
  Department of Materials Physics, Graduate School of Engineering, Nagoya University, Furo-cho, Chikusa, Nagoya 464-8603, Japan
}%



\date{\today}

\begin{abstract}
Complex systems such as protein conformational fluctuations and supercooled liquids  exhibit a long relaxation time 
and are considered to posses multiple relaxation times. We analytically obtain the exact correlation function for stochastic processes with multiple relaxation times. 
We show that the time-averaged correlation function calculated by a trajectory whose length is shorter than the longest relaxation time exhibits an apparent aging behavior. 
We propose a method to extract relaxation times from a single trajectory. This method successfully extracts relaxation times of a stochastic process with multiple states 
when a state can be characterized by the values of the trajectory. 
As an application of this method, we estimate several relaxation times smaller than the longest relaxation time in conformational fluctuations of a small protein. 
\end{abstract}

\maketitle


\section{Introduction}

A power-law decay in the correlation function or the power spectral density (PSD), i.e., $1/f$ noise, 
is one of the typical features in complex systems such as disordered systems 
\cite{Scher1975, Monthus1996, kuno2000nonexponential, frantsuzov2008universal} and biological systems \cite{Peng1994, yang2003protein, Wong2004, Weigel2011, Yamamoto2013, Yamamoto2014b,yamamoto2015origin,hu2016dynamics}. 
In many cases, there is an exponential cutoff in the correlation because of the finite size effect or the finite time scale. 
A superposition of multiple exponential relaxation modes is one of the origins of a power-law decay of 
the correlation function \cite{Richert-2002}. 
The Rouse and Zimm models for polymers exhibit such a power-law decay
behavior for the stress relaxation functions \cite{Doi-Edwards-book}.
In these models, a polymer chain is modeled by connecting beads with harmonic springs. Due to the connectivity, the motions of beads are
correlated and we observe multiple relaxation modes. On a short-time
scale, the superposition of relaxation modes gives a power-law type relaxation.
For a long timescale, only the relaxation time with the longest relaxation
time survives and the decay becomes exponential.
Such a power-law decay with an exponential cutoff is often 
observed in the Ornstein-Uhlenbeck process with fluctuating diffusivity \cite{Uneyama2019} as well 
as protein conformations \cite{Yamamoto2014b, hu2016dynamics, Yamamoto2021}. 
It is also considered  that several exponential relaxation modes are superposed in supercooled liquids \cite{Richert-2002}. 

Extracting relaxation modes from a random signal is an important subject especially when 
the random signal is composed of several states.
If we can obtain a sufficient number of samples and have the accurate
correlation function, we can directly decompose the correlation function
into relaxation modes by utilizing some numerical methods \cite{Brezinski-book}.
The decomposition can be done manually (so-called the Procedure X) \cite{Tobolsky-Murakami-1959}.
If we have the information on trajectories (which can be easily
obtained in molecular simulations), we can utilize some elaborated methods
such as the relaxation mode analysis \cite{takano1995relaxation, hagita2002relaxation,mitsutake2011relaxation}.
Moreover, a time-structure-based independent component analysis also reveals several relaxation modes in protein fluctuations 
\cite{naritomi2011slow}.
For the analysis methods shown above, a suffiiciently large number
of statistically independent samples and/or detailed trajectory data are required. 
However, in experiments, 
the data is limited and sometimes low dimensional.
In molecular simulations, the data are sometimes also limited.
For example, the relaxation modulus can 
be calculated as the correlation function of the shear stress,
by utilizing the linear response theory \cite{Evans-Morris-book}.
The stress tensor has only $6$ components and thus the number of data
points is limited in many practical cases. 
Therefore, a method to extract the exact relaxation modes from low-dimensional data is necessary for data analysis.   

A dichotomous process, where a random signal $I(t)$ is composed of two states, is one of the simplest stochastic models exhibiting a power-law decay of the correlation function 
\cite{God2001, burov2010aging, Niemann2013, Leibovich2015}. When the probability density function (PDF) of the duration time for 
one of the states follows a 
power-law distribution, the correlation function decays a power law. Moreover, the correlation function $\langle I(t) I(t+\Delta) \rangle$ 
depends on time $t$, i.e., aging, when the mean duration time diverges. Such an aging behavior is a signature of non-stationary processes. 
Aging can also be detected in time-averaged quantities defined by $\overline{\mathcal O} (T) \equiv \int_0^T {\mathcal O} (t')dt'/T$. 
In aging systems, a time-averaged quantity explicitly depends on $T$. In particular, the diffusion coefficient defined 
by the time-averaged square displacement depends on measurement time in stochastic models of anomalous diffusion 
\cite{He2008, Neusius2009, Miyaguchi2011, Miyaguchi2013, Akimoto2013a, Akimoto2014, Metzler2014, Miyaguchi2015}.
However, an explicit dependence of the measurement time in the correlation function sometimes appears for stationary processes. 
For example, the time-averaged correlation of the inter-domain distance depends on the measurement time 
 whereas its PSD does not depend on the measurement time \cite{hu2016dynamics}. This apparent aging behavior is still controversial \cite{goychuk2021insufficient}.

This paper aims at obtaining different relaxation times from a single trajectory. 
To this end, we consider a stationary stochastic process that has multiple relaxation times. 
In particular, we consider a superposition of dichotomous processes and a superposition of Ornstein-Uhlenbeck processes. 
We obtain the exact form of the correlation functions for these models. Furthermore, 
we find that 
these models show apparent aging behaviors in the time-averaged correlation functions when the measurement time is smaller than the longest relaxation time. 
When the time scales of relaxation times and the magnitudes of the state values are separable, we obtain several relaxation times from the single trajectory. 
We apply this method to protein conformational fluctuations and estimate several relaxation times. 

\section{Stochastic processes with multiple relaxation times}

\subsection{Superposition of dichotomous processes}

Here, we consider a multi-state ($2N$-state) stochastic process, where $N$ dichotomous processes with different relaxation times 
$\tau_k$ are superposed ($k=1, \cdots, N$). In each dichotomous process, the random signal takes two values, i.e., $I_k^+$ or $I_k^-$ 
for the $k$-th dichotomous process ($k=1, \cdots, N$). We assume that changes in the random signal are stochastic.
In the  superposed process, the resulting random signal $I(t)$ at time $t$ 
is represented by
\begin{equation}
I(t) = \sum_{k=1}^N I_k(t), 
\end{equation}
where $I_k(t)$ is the random signal of the $k$-th dichotomous process at time $t$. 
In each dichotomous process, the duration time of each state is an independent and identically distributed (IID) random variable, and 
the PDF for the $k$-th dichotomous process is given by $\psi_k(t)$. Here, we assume that 
the PDFs for both states, i.e., $I_k^+$ and $I_k^-$, for dichotomous process $I_k(t)$ are identical for simplicity. We will discuss a general case later. 
We note that all dichotomous processes are independent of each other. 

First, let us derive the correlation function for each dichotomous process. The normalized correlation function is defined as 
\begin{equation}
C(t) = \frac{\langle I(t) I(0) \rangle - \langle I(0) \rangle^2}{\langle I(0)^2 \rangle - \langle I(0) \rangle^2},
\end{equation}
where $\langle \cdot \rangle$ represents the ensemble average. By definition, $C(0)=1$. 
In the following, we assume the stationarity. Under the assumption, 
$\langle I(t)^2 \rangle$ does not depend on $t$, i.e., $\langle I(t)^2 \rangle=\langle I(0)^2 \rangle$ for all $t$. 
Furthermore, $\langle I(t+t') I(t') \rangle$ does not depend 
on $t'$. 
Let $N_t$ be the number of stochastic changes of $I_k(t)$ until $t$ and $\chi_n$ be the $n$-th state value, i.e., $\chi_n=I_k^+$ or 
$I_k^-$. The unnormalized correlation function 
for $k$-th dichotomous process, i.e., $\hat{C}_k(t)=\langle I_k(t) I_k(0) \rangle - \langle I_k(0) \rangle^2 $, 
 can be calculated as
\begin{eqnarray}
\hat{C}_k(t) 
&=& \Pr (N_t =0) \langle \chi_0^2 \rangle + \sum_{n=1}^\infty \Pr (N_t =n) \langle \chi_0 \chi_n \rangle - \langle I_k \rangle^2
\nonumber\\
&=& (\langle I_k^2 \rangle  - \langle I_k \rangle^2 ) \Pr (N_t =0), 
\end{eqnarray}
where $\langle I_k \rangle = \langle I_k(0) \rangle $, $\langle I_k^2 \rangle = \langle I_k(0)^2 \rangle $, and
 we use $\langle \chi_0 \chi_n \rangle = \langle \chi_0 \rangle \langle \chi_n \rangle$ because $\chi_0$ and $\chi_n$ are independent. 
 When the first waiting time distribution is identical to $\psi_k(t)$, i.e., {\it ordinary renewal process} \cite{Cox}, the probability that there are no stochastic changes until time $t$ 
 becomes
  \begin{equation}
  \Pr (N_t =0) =\int_t^\infty \psi_k(x) dx.
  \end{equation}
 In particular, when the PDF follows an  exponential distribution, i.e., $\psi_k(t) = \tau_k^{-1} \exp (- t/\tau_k)$, the correlation function becomes
 \begin{equation}
 C_k(t) =  
 \exp( -t/\tau_k). 
 \end{equation}
 On the other hand, when the PDF follows a power-law distribution, i.e., $\psi_k(t) = \alpha t^{-1-\alpha}$ ($\alpha>1$) for $t\geq 1$ and 0 otherwise, 
the correlation function exhibits a power-law decay:
\begin{eqnarray}
 C_k(t) =  \left\{
 \begin{array}{ll}
 1 &(t<1)\\
 \\
  t^{-\alpha}\quad &(t\geq 1).
  \end{array}
\right.   
 \end{eqnarray}
Note that the condition of $\alpha>1$ is necessary for stationarity. The mean duration time diverges for $\alpha \leq 1$, which 
exhibits non-stationary and aging behaviors. Furthermore, the correlation function strongly depends on 
the first duration time distribution for $\alpha<2$ \cite{Akimoto2007}. In {\it equilibrium renewal processes} \cite{Cox,Miyaguchi2013,Akimoto2018b}, 
where the PDF of the first duration time is not $\psi_k(t)$ but the equilibrium distribution, i.e., $\int_t^\infty \psi_k(t')dt'/\int_0^\infty \psi_k(t')dt'$, 
the correlation function becomes 
\begin{eqnarray}
 C_k(t) =  \left\{
 \begin{array}{ll}
 1- (1-\frac{1}{\alpha})t &(t<1)\\
 \\
 \frac{1}{\alpha} t^{-\alpha+1}\quad &(t\geq 1).
  \end{array}
\right.   
 \end{eqnarray}
 Therefore, the power-law exponent becomes different from that in the ordinary renewal process. 
Such an initial ensemble dependence is often observed in  anomalous diffusion 
\cite{Akimoto2012, Akimoto2018b}.

Next, we consider the superposed process. The unnormalized correlation function 
is given by the sum of $\hat{C}_k(t)$:
\begin{eqnarray}
\hat{C}(t) 
&=& \sum_{k=1}^N (\langle I_k^2 \rangle  - \langle I_k \rangle^2 ) \int_t^\infty \psi_k(x) dx ,
\end{eqnarray}
where we assume that the first waiting time distribution for $I_k(t)$ is $\psi_k(t)$, i.e., {\it ordinary renewal process} \cite{Cox}.
If the duration time PDFs of all the dichotomous processes follow the exponential distribution with relaxation times $\tau_k$ 
($k=1, \cdots, N$), the correlation function shows a superposition of the exponential decays: 
\begin{equation} 
C(t) = \sum_{k=1}^N\frac{ (\langle I_k^2 \rangle  - \langle I_k \rangle^2 ) }{\langle I(0)^2 \rangle - \langle I(0) \rangle^2} \exp(-t/\tau_k).
\label{correlation-multi-exp}
\end{equation}

\subsection{Multi-state process with multiple transitions}
Here, we derive the correlation function of another multi-state process, where 
there are $N$ states and the state values take $I_k$ ($k=1, \cdots, N$). 
We assume that the probability of transition from state $i$ to $j$ is given by $p_{ij} = p_j$ and that the probability of the initial 
state being $j$ is also given by $p_j$. By the same calculation as in the above, we have
\begin{eqnarray}
\hat{C}(t) 
&=& \sum_{k=1}^N p_k ( I_k^2   - I_k \langle I (0) \rangle )\int_t^\infty \psi_k(x) dx ,
\end{eqnarray}
where $\langle I(0) \rangle=\sum_{k=1}^N p_k I_k$ and we assume the ordinary renewal process. 
If the duration time PDFs of all the dichotomous processes follow the exponential distribution with relaxation times $\tau_k$ 
($k=1, \cdots, N$), the correlation function decays  as
\begin{equation} 
C(t) = \sum_{k=1}^N  \frac{p_k I_k ( I_k - \langle I (0) \rangle )}{\langle I(0)^2 \rangle - \langle I(0) \rangle^2} \exp(-t/\tau_k).
\end{equation}
Although the correlation function is the same form as Eq.~\eqref{correlation-multi-exp}, the process is not a superposition 
of dichotomous processes.

\subsection{Superposition of alternating renewal processes}

An alternating renewal process with states  $I_+$ and $I_-$ is a dichotomous process with different duration time PDFs \cite{Cox}. 
The duration time PDFs for $I_+$ and $I_-$ are  different, i.e., $\psi^+(t)$ and $\psi^+(t)$, respectively. Transitions from $I_+$ to $I_-$ 
or vice versa are deterministic. The probability of finding a $+$ state at time $t$ when $I(0)=I_+$ is given by
\begin{equation}
p_+(t) = \sum_{k=0}^{\infty} \Pr (N_t = 2k).
\end{equation}
By the same calculation as in Ref.~\cite{Miyaguchi2016}, in the long-time limit, the probability does not depend on the initial condition. 
Therefore, the equilibrium probabilities of finding $+$ and $-$ states are given by
\begin{equation}
p_{\pm}^{\rm eq} = \frac{\tau^{\pm}}{\tau^+ + \tau^-},
\end{equation}
where $\tau^+$ and $\tau^-$ are the mean duration times for  $+$ and $-$ states, respectively. We assume that they are finite.

We consider the correlation function for the alternating renewal process, where the initial condition is the equilibrium one. 
In equilibrium, 
\begin{equation}
\langle I(0) \rangle = p_+^{\rm eq} I_+ + p_-^{\rm eq}I_-
\end{equation}
and 
\begin{equation}
\langle I(0)^2 \rangle = p_+^{\rm eq} I_+^2 + p_-^{\rm eq} I_-^2.
\end{equation}
Moreover, we have
\begin{eqnarray}
\langle I(t)I(0) \rangle &=&  W_{++}^{\rm eq} (t) I_+^2 + W_{+-}^{\rm eq} (t) I_+ I_{-}  + W_{-+}^{\rm eq} (t) I_- I_{+}\nonumber\\
&&+ W_{--}^{\rm eq} (t) I_-^2,
\end{eqnarray}
where $W_{hh'}^{\rm eq}(t)$ is  the transition probabilities from $h$ ($=+$ or $-$) to $h'$ ($=+$ or $-$) at time $t$.   The Laplace transform of 
$W_{hh'}^{\rm eq}(t)$ with respect to $t$ is obtained as \cite{Miyaguchi2016}
\begin{equation}
\hat{W}_{\pm \pm}^{\rm eq} (s) = \frac{ p_{\pm}^{\rm eq}}{s}  - \frac{1}{\mu s^2} \frac{[1- \hat{\psi}_+(s)][1- \hat{\psi}_-(s)]}{1 - \hat{\psi}(s)},
\end{equation}
and
\begin{equation}
\hat{W}_{\pm \mp}^{\rm eq} (s) = \frac{1}{\mu s^2} \frac{[1- \hat{\psi}_+(s)][1- \hat{\psi}_-(s)]}{1 - \hat{\psi}(s)},
\end{equation}
where $\mu =\tau^+ + \tau^-$ and $\hat{\psi}(s) = \hat{\psi}_+(s) \hat{\psi}_-(s)$. When the PDFs $\psi_+(t)$ and $\psi_-(t)$ follow 
exponential distributions, the transition probabilities become
\begin{equation}
W_{+ -}^{\rm eq} (t) = p_+^{\rm eq} p_-^{\rm eq} (1- e^{-\frac{\tau^+\tau^-}{\mu}t}).
\end{equation}
Therefore, the correlation function decays as
\begin{equation}
\langle I(t)I(0) \rangle  - \langle I(0) \rangle^2  = p_+^{\rm eq} p_-^{\rm eq} (I_+ - I_-)^2 e^{-\frac{\tau^+\tau^-}{\mu}t}.
\end{equation}

We consider a superposition of alternating processes, i.e., $I(t) = \sum_{k=1}^N I_k(t)$. We assume that alternating renewal processes, i.e.,  
$I_1(t), I_2(t), \cdots, I_N(t)$, are independent and the duration time PDFs for $I_{k}^{\pm}$, which is the $\pm$ state for $I_k(t)$, follow exponential
 distributions with mean $\tau_k^\pm$.   
  Therefore, the correlation function can be obtained as the sum of the correlation functions for $I_k$. It follows that 
 the correlation function becomes 
 \begin{equation}
C(t) = \sum_{k=1}^N 
\frac{p_{k,+}^{\rm eq} p_{k,-}^{\rm eq} (I_{k}^{+} - I_{k}^{-})^2}{\langle I(0)^2 \rangle  - \langle I(0) \rangle^2 } e^{-\frac{\tau_{k}^{+}\tau_{k}^{-}}{\mu_k}t},
\end{equation}
where $\mu_k = \tau_k^+ + \tau_k^-$ and $p_{k,\pm}^{\rm eq} = \tau_k^\pm/\mu_k$.

\subsection{Superposition of Ornstein-Uhlenbeck processes}

Here, we we derive the correlation function of Ornstein-Uhlenbeck (OU) processes. 
Dynamic equation for the OU process is given by 
\begin{equation}
 \label{ou-multi}
 \frac{dI_i (t)}{dt} = - \nu_i I_i(t) 
  + \sqrt{2 \nu_i \eta_i} \xi_i (t),
\end{equation}
where $\nu_i$ is constant and $\xi_i(t)$ is a white Gaussian noise, i.e., 
$\langle \xi_{i}(t) \rangle =0$ and $\langle \xi_i (t) \xi_j(t') \rangle = \delta_{ij} \delta(t-t')$.
The autocorrelation function of this process shows an exponential relaxation:
\begin{equation}
 \hat{C}_{i}(t) = \langle I_i(0) I_i(t) \rangle = 
  \eta_i \exp(-\nu_i t).
\end{equation}
For a superposition of the OU processes, the random signal is given by $I(t) = \sum_{i=1}^N I_i(t)$. Similar to a superposition 
of dichotomous processes, the correlation function can be represented by the sum of those for the OU processes, i.e., 
$\hat{C}(t) = \sum_{i=1}^N \hat{C}_i(t)$. 
Therefore, the correlation function of the superposed 
process becomes
\begin{equation}
C(t) = \frac{\sum_{i=1}^N \eta_i \exp(-\nu_i t)}{\sum_{k=1}^N \eta_i  }.
\label{ou-theory}
\end{equation}

\section{Method to extract relaxation times from a single trajectory}

\subsection{Apparent aging behavior}
In some experiments, it is difficult to obtain a plenty number of trajectories. In these cases, the number of the ensemble is not enough
 to calculate the exact ensemble average. 
Therefore, instead of the ensemble average, the time average is used to calculate the correlation function \cite{Brokmann2003,  Golding2006, Weigel2011}. 
Using time averages
\begin{equation}
\overline{I} (T) = \frac{1}{T} \int_0^{T} I(t') dt' ,
\end{equation}
and 
\begin{equation}
\overline{\hat{C}}(t;T) = \frac{1}{T-t} \int_0^{T-t} I(t'+t)I(t') dt' -\overline{I} (T)^2,
\label{ta-correlation}
\end{equation}
where $T$ is the measurement time, 
we have the time-averaged correlation function: 
\begin{equation}
\overline{C} (t; T) = \frac{\overline{\hat{C}}(t;T)}{\overline{\hat{C}}(0;T)}.
\label{ta-cor}
\end{equation}
For stationary ergodic stochastic processes, the time-averaged quantity coincides with the corresponding ensemble 
average in the long-time limit. Here, we assume the ergodic property:
\begin{equation}
\overline{C} (t; T)  \to C (t) \quad (T\to\infty).
\end{equation}
Although the above assumption is considered to be valid in many experiments, 
the breaking of ergodicity is substantially studied in anomalous diffusion \cite{He2008, Neusius2009, Miyaguchi2011, Miyaguchi2013, Akimoto2013a, Akimoto2014, Metzler2014, Miyaguchi2015} or blinking quantum dots \cite{Brokmann2003}, where 
the time-averaged quantities such as the diffusion coefficient and the blinking ratio do not converge to a constant but 
remain random. In this situation, the decay of the time-averaged correlation function differs from that of the ensemble-averaged correlation function \cite{burov2010aging}. 

\begin{figure}
\includegraphics[width=.9\linewidth, angle=0]{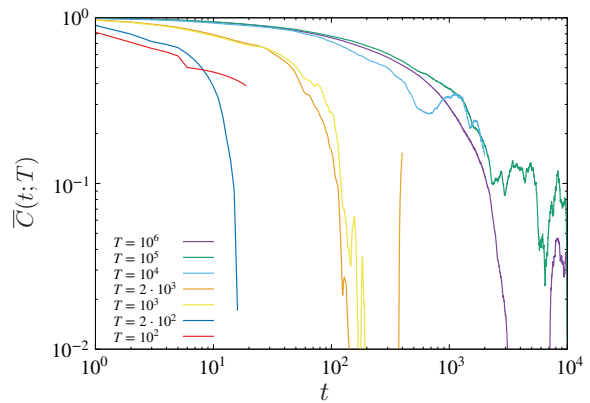}
\caption{Time-averaged correlation functions with different measurement times ($T=10^2, 2\cdot 10^2, 10^3, 2\cdot 10^3, 10^4, 10^5$, and $10^6$) 
for a single trajectory, where three 
dichotomous processes with relaxation times are superposed ($\tau_k=10,10^2$, and $10^3$ for $k=1,2,$ and 3, respectively). 
The state values are $I_1^\pm=\pm 0.25$, $I_2^\pm=\pm 0.5$, and $I_3^\pm=\pm 1$.}
\label{apparent-aging-cor}
\end{figure}

Even in a stationary ergodic stochastic process, the time-averaged correlation function  may differ 
from the ensemble-averaged correlation function
when the measurement time $T$ is much smaller than the longest relaxation time. 
Furthermore, we find that the time-averaged correlation function 
exhibits an {\it apparent aging behavior} in a multi-state stochastic process such as a superposition of dichotomous processes 
(see Fig.~\ref{apparent-aging-cor}). The time-averaged correlation functions become different 
in different measurement times. 
This apparent aging behavior is also observed for time series with a single relaxation mode when the relaxation time 
is greater than the measurement times (see Appendix.~A). 
Thus, the time-averaged correlation function sometimes fails to capture the relaxation time. 

\subsection{Freezings of states with longer relaxation times}
Here, we explain the reason why the time-averaged correlation functions show apparent aging behaviors. 
For simplicity, we consider a superposition of two dichotomous processes, 
where the relaxation time for one of the dichotomous processes $\tau_1$ is much larger than measurement time $T$. 
Because the state value with the longer relaxation time rarely changes during the measurement, 
the state value with the longer relaxation time  is almost constant, e.g., $I_1(t) = I_1^+$ or 
$I_1^-$ for $t<T$. We call it {\it freezing} of the state. Therefore, the state value can be 
approximately decomposed as $I(t) = I_1^+ + I_2(t)$ or $I_1^- + I_2(t)$. When $I(t) = I_1^+ + I_2(t)$ for $t<T$, 
 the time-averaged unnormalized correlation function is approximately calculated as
\begin{eqnarray}
\overline{\hat{C}} (t; T) &\cong& \frac{1}{T-t} \int_0^{T-t} I_2(t'+t)I_2(t') dt'  + {I_1^+}^2\nonumber\\
&& + 2 I_1^+ \overline{I_2}(T) - \{ I_1^+ + \overline{I_2}(T)\}^2.
\end{eqnarray}
It follows that the time-averaged correlation function becomes 
\begin{eqnarray}
\overline{C} (t; T) \cong C_2(t) , 
\end{eqnarray}
where the relaxation time of $I_2(t)$ is much smaller than $T$. Although the time-averaged correlation function 
 fails to capture the exact relaxation time, one can estimate the relaxation time $\tau_2$ from the time series.

A similar argument holds for the process that is a superposition of $N$ dichotomous processes, where the freezings of states 
are observed for states whose relaxation times are longer than the measurement time. Therefore, the time-averaged correlation 
functions will depend on the measurement times. In particular, when the measurement time $T$ is smaller than $\tau_{k+1}$, which is 
the $(k+1)$-th relaxation time, i.e., 
$\tau_1 < \tau_2 < \cdots < \tau_k < \cdots <\tau_N$, the time-averaged correlation function of a trajectory $I(t)$ with $t<T$ 
and $\tau_k <T<\tau_{k+1}$ becomes
\begin{equation}
\overline{C} (t; T) \cong \sum_{i=1}^k C_i(t).
\end{equation}
This result suggests that multiple relaxation times can be extracted by changing the 
measurement time if the process is a superposition of multiple dichotomous processes and the orders of the relaxation times are separated.

\subsection{ Extracting relaxation times from a single trajectory}
\begin{figure}
\includegraphics[width=.9\linewidth, angle=0]{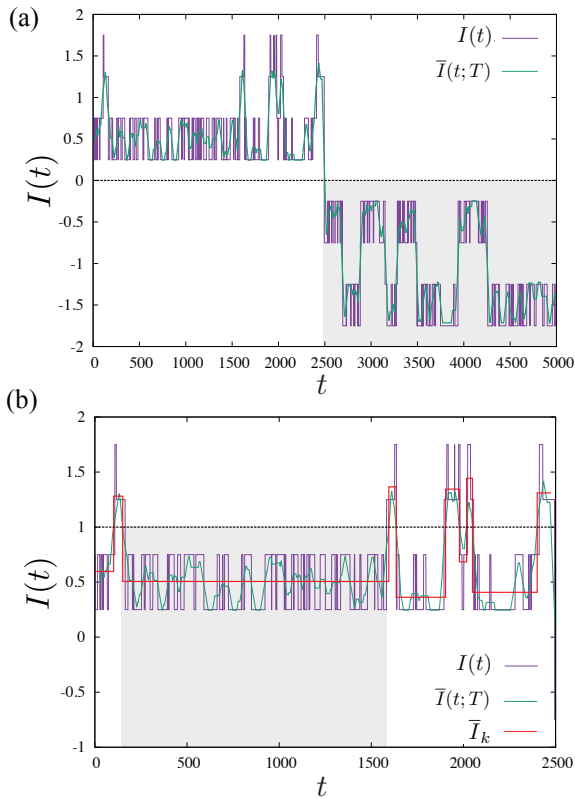}
\caption{Trajectory and the corresponding widow time average of $I(t)$ for a superposition of three dichotomous processes,
where the duration-time distributions are the exponential distribution with means $\tau_1=10^3$, $\tau_2=10^2$, and $\tau_3=10$ and 
the state values are $I_1^\pm=\pm 1$, $I_2^\pm=\pm 0.5$, and $I_3^\pm=\pm 0.25$. (a)The window time average is calculated by $T=50$.
(b) Amplification of (a), where time average $\overline{I}_k $ is added. The time-averaged correlation functions are calculated in the shaded areas.}
\label{traj-tasv}
\end{figure}

{\color{black}
To extract relaxation times, we need to know the transition points of states. Several methods were proposed to unveil transition points 
\cite{montiel2006quantitative, Akimoto2017}.
According to Ref.~\cite{Akimoto2017}, one can estimate transition points of diffusive states \cite{Yamamoto2021}. 
Here, we apply this method to state values. We assume that time scales of relaxation times are separable, i.e., $\tau_1 \gg \tau_2 \gg \cdots$. 
We use a window time average of $I(t)$ as a function of $t$, i.e., 
\begin{equation}
\overline{I}(t;T_w) \equiv \frac{1}{T_w} \int_{t-T_w/2}^{t+T_w/2} I(t')dt',
\end{equation}
where $T_w$ is the length of the window time average, which is a tuning parameter. When a duration time of a state is longer than $T_w$, a transition point of a state 
can be obtained by a point at which 
$\overline{I}(t;T_w)$ changes extensively. First, we determine a threshold value $I_{\rm th}$, which is defined as $I_{\rm th} = (I_{\max} + I_{\min})/2$, where 
$I_{\max} = \max \{ I(t) | t \in [0,T] \}$, $I_{\min} = \min \{ I(t) | t \in [0,T] \}$, and $T$ is the total measurement time. 
Points at which $\overline{I}(t;T_w)$ crosses $I_{\rm th}$ are candidates for the transition points. 
Figure~\ref{traj-tasv}(a) shows a trajectory and a window time average of $I(t)$ for a superposition of three dichotomous processes.  
The state value changes extensively around 2500. The crossing point $t_c$ corresponds to a transition point. The state values changes 
around $t_c$. 
For $t<t_c$, we redefine the threshold value by $I_{\rm th} = (I_{\max} + I_{\min})/2$ in the time interval, i.e., $[0,t_c]$. 
By a statistical test which is the same as in Ref.~\cite{Akimoto2017}, 
 we have correct transition points $t_1, t_2, \cdots$ with $t_1 < t_2 < \cdots$.  
 We have time averages of $I(t)$ in the interval $[t_k, t_{k+1})$:
  \begin{equation}
\overline{I}_k  \equiv  \frac{1}{t_{k+1}-t_k } \int_{t_k}^{t_{k+1}} I(t')dt' .
\end{equation}
Figure~\ref{traj-tasv}(b) shows $I(t), \overline{I}(t;T)$, and $\overline{I}_k $. For time interval $[t_2, t_3]$ in Fig.~\ref{traj-tasv}(b), one of states does not changes, i.e., freezing occurs.
We obtain the time-averaged correlation 
function in the time interval $[t_k, t_{k+1}]$:
\begin{eqnarray}
\overline{C}_{(k)} (t; T_k) \equiv 
&&\dfrac{1}{t_{k+1}-t_k -t} \int_{t_k}^{t_{k+1}-t} I(t'+t)I(t')dt'  \nonumber\\
&&- \overline{I}_k ^2,
\end{eqnarray}
where $T_k = t_{k+1} - t_{k}$.
Using the time-averaged correlation function $\overline{C}_{(k)} (t)$, one can estimate relaxation times. 

We test this method to time series of a superposition of three dichotomous processes with relaxation times $\tau_1=10^3$, $\tau_2=10^2$, and $\tau_3=10$.
The longest relaxation time can be obtained by the time-averaged correlation function when the measurement time is much longer than the longest relaxation time. 
Other relaxation times smaller than the longest relaxation time can be obtained by the time-averaged correlation function of a time series if a freezing occurs in the 
time interval.
As shown in Fig.~\ref{traj-tasv}, several freezings occur in the time series.  We obtain a relaxation time as $\tau\cong 8.3$ 
from $\overline{C}_{(2)} (t; 1444)$ in  Fig.~\ref{traj-tasv}(b), i.e., the time-averaged correlation function in the interval $[t_2, t_{3})$. 
This relaxation time corresponds to $\tau_3=10$. 
Moreover, we also estimate a relaxation time as $\tau\cong 10^2$ from $\overline{C}_{(1)} (t; 2506)$ in  Fig.~\ref{traj-tasv}(a), 
 i.e., the time-averaged correlation function in the interval $[t_1, t_{2})$. This relaxation time corresponds to $\tau_2=10^2$. 
 Therefore, we estimate it successfully. 
We note that trajectories used in the above estimations are different.}
  
\begin{figure*}
\includegraphics[width=.9\linewidth, angle=0]{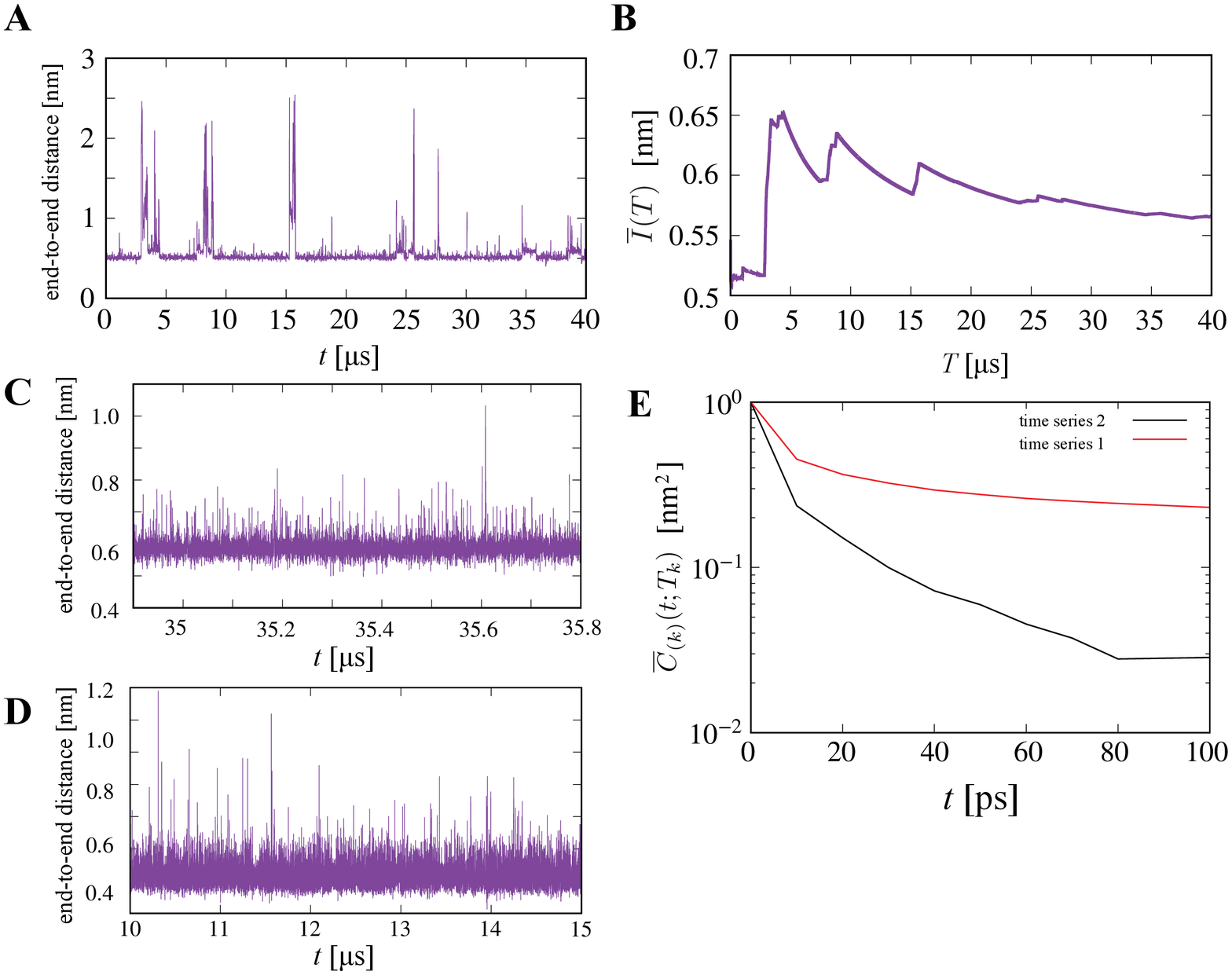}
\caption{(A) Time series of the end-to-end distance between C${}_\alpha$ atoms of small proteins, where data are plotted every 10 ns.
(B) Time average of end-to-end distance $I(t)$ as a function of the measurement time $T$.  
(C)  Blowup of the time series of (A) (Time series 1).
(D)  Blowup of the time series of (A) (Time series 2).
(E) Time-averaged correlation functions for time series 1 and 2. 
}
\label{protein-apparent-cor}
\end{figure*}

\section{Application to protein fluctuations}

Here, we apply our method to protein conformational fluctuations. We estimate relaxation times of the end-to-end 
distance between C${}_\alpha$ atoms of a small protein, super Chignolin \cite{honda2008crystal}, isolated in solution. 
In what follows, we denote the end-to-end distance by $I(t)$. 
The data were produced by molecular dynamics simulation. 
The details of the simulation method and conditions are described in Ref.~\cite{Yamamoto2021}. 
We have a time series of the end-to-end distance with a length of 40 \textmu{}s. 
In the previous study \cite{Yamamoto2021}, the longest relaxation time is estimated as around 1 \textmu{}s. 

Here, we estimate relaxation times shorter than the longest relaxation time for a single trajectory of the protein conformational fluctuations.
Figure~\ref{protein-apparent-cor} shows the time series of the end-to-end distance, the time average of  the end-to-end distance, 
and the correlation functions. 
As shown in Fig.~\ref{protein-apparent-cor}B, the time average $\overline{I}(T)$ converges to a value for large $T$ and $\overline{I}(T)$ 
for $T < 2$ \textmu{}s is smaller than the convergence value. This implies that we have to use a long trajectory to obtain the exact correlation function.
We divide the time series into different states using the above method.  Figures \ref{protein-apparent-cor}C and \ref{protein-apparent-cor}D 
show the time series that we use to extract the relaxation times. In particular, we calculate the time-averaged correlation functions for 
a time series  with a time interval from 10  \textmu{}s to 15  \textmu{}s  (time series 1) and a time series  around 35  \textmu{}s (time series 2).
Figure~\ref{protein-apparent-cor}D shows the time-averaged correlation functions for the two time series. The relaxation times are clearly 
different from each other and much smaller than the longest relaxation time. Although the correlation functions do not decay exponentially, 
we roughly estimate the relaxation times for time series 1 and 2  by 40 ps and 20 ps, respectively. 
This result suggests that multiple relaxation times are superposed in the protein conformal fluctuations. 
Therefore, $1/f$ noise widely observed in protein conformational fluctuations originates from a superposition of multiple relaxation modes.

\section{Conclusion}

We derive the exact forms of the correlation function in multi-state stochastic processes. When the length of a trajectory is 
smaller than a relaxation time, the time-averaged correlation function fails to capture the relaxation time. In other words, 
an apparent aging behavior is observed due to the freezing of a state. In this paper, we have proposed a method to extract 
relaxation times from a single trajectory even when multiple relaxation times are embedded. 
This method is successfully applied to time series with multiple relaxation times when the value of the time series corresponds to the state. 
Although we assume stationarity of time series in this paper, 
we expect that this method to extract relaxation times 
can be applied to non-stationary time series because we use a time series whose length is less than the longest relaxation time.
Our method can be applied to estimate relaxation times of a single trajectory. We expect that if we calculate the self intermediate
scattering function from a single particle trajectory, we can apply our method to estimate relaxation times of the single particle dynamics. 
Although the estimated relaxation times are relevant to relaxation times for a macroscopic system, 
the connection is not simple in many-body complex systems such as supercooled liquids.

Finally, we discuss apparent aging in a stationary time series. 
When the orders of multiple relaxation times are different, {\it freezings} of states 
will occur. We have shown that the time-averaged correlation function depends on the window size $T$. 
 In other words, the time-averaged correlation function exhibits an apparent aging behavior. 
Furthermore, we also find that such an apparent aging behavior is observed when a trajectory is much smaller than the relaxation time. 
Therefore, an apparent aging behavior observed in protein conformational fluctuations will originate from an insufficient length of the time series. 

\section*{Acknowledgement}
T.A. was supported by JSPS Grant-in-Aid for Scientific Research (No.~C JP18K03468). E.Y. was partially supported by JSPS Grant-in-Aid for Scientific Research (No. 20K14432).
T.U. was supported by Grant-in-Aid for Scientific Research Grant B (No.~JP19H01861),
Grant-in-Aid (KAKENHI) for Transformative Research Areas B (No.~JP20H05736),
and JST, PRESTO (No.~JPMJPR1992).

\appendix

\section{Derivation of time-averaged correlation function for Orstein-Uhlenbeck process}
Here, we derive the time-averaged correlation function, defined by Eq.~\eqref{ta-correlation}, for a one-dimensional OU process. 
The OU process is a simple Gaussian process and the ensemble-averaged correlation function becomes a single exponential form.
We consider a single-mode OU process with $N = 1$, $\nu_{1} = 1$, and $\eta_{1} = 1$ in Eq.~\eqref{ou-multi}.
The OU process for $I(t)$ becomes 
\begin{equation}
 \label{ou-single}
 \frac{dI(t)}{dt} = - I(t) + \sqrt{2} \xi(t),
\end{equation}
where $\xi(t)$ is the Gaussian noise which satisfies
$ \langle \xi(t) \rangle = 0$
and $\langle \xi(t) \xi(t') \rangle = \delta(t - t')$.
The solution of the OU process \eqref{ou-single} is
\begin{equation}
 I(t) = I(0) e^{-t} + \sqrt{2} \int_{0}^{t} ds \, e^{-(t - s)} \xi(s),
\end{equation}
and the ensemble-averaged correlation function is calculated to be
\begin{equation}
\begin{split}
   \langle I(t) I(0) \rangle
 & = \langle I^{2}(0) \rangle e^{-t} 
  + \sqrt{2} \int_{0}^{t} ds \, e^{-(t - s)} \langle I(0) \xi(s) \rangle \\
 & = e^{-t} ,
\end{split}
\end{equation}
where the initial ensemble is the equilibrium distribution, i.e., $\langle I(0) \rangle=0$ and $\langle I^2(0) \rangle=1$.

As shown in the main text, if we subtract the time average of $I(t)$
from $I(t)$, the correlation function apparently changes. The time average
for a finite measurement time $\bar{I}(T)$ is generally non-zero whereas
the ensemble average of $\langle I \rangle$ is zero.
The time-averaged correlation function for $I(t) - \bar{I}(T)$ can be represented as 
\begin{equation}
 \label{ou-correlation-subtraction-definition}
 C'(t;T) = \frac{1}{T - t}
  \int_{0}^{T - t} dt' [I(t') - \bar{I}(T)]
  [I(t' + t) - \bar{I}(T)] ,
\end{equation}
where $\bar{I}(T)$ is the time-average of $I(t)$ over a finite measurement
time:
\begin{equation}
 \bar{I}(T) \equiv \frac{1}{T} \int_{0}^{T} dt' \, I(t').
\end{equation}
$C'(t;T)$ defined by Eq.~\eqref{ou-correlation-subtraction-definition}
does not coincide to the ensemble-averaged correlation function except
the limit of $T \to \infty$.
Using the correlation function for $I(t)$ without the subtraction,
\begin{equation}
 C(t;T) \equiv \frac{1}{T - t} \int_{0}^{T - t} dt' \,
  I(t') I(t' + t), 
\end{equation}
we have
\begin{widetext}
\begin{equation}
 \label{ou-correlation-subtraction}
 \begin{split}
  C'(t;T) 
  & = C(t; T)   + \bar{I}^{2}(T)
  - \frac{1}{T - t} \int_{0}^{T - t} dt' \,    \bar{I}(T) I(t')
  - \frac{1}{T- t} \int_{t}^{T} dt' \,  \bar{I}(T) I(t') \\
  & = C(t;T)   + \bar{I}^{2}(T)
  - \frac{T}{T - t} \frac{1}{T} \int_{0}^{T} dt' \,    \bar{I}(T) I(t')
  + \frac{1}{T - t} \int_{T - t}^{T} dt' \,    \bar{I}(T) I(t') \\
  & \qquad - \frac{T}{T - t} \frac{1}{T}\int_{0}^{T} dt' \,  \bar{I}(T) I(t') 
  +  \frac{1}{T - t} \int_{0}^{t} dt' \,  \bar{I}(T) I(t') \\
  & = C(t;T) - \frac{T + t}{T - t}\bar{I}^{2}(T)
  + \frac{1}{T - t} \int_{T - t}^{T} dt' \,    \bar{I}(T) I(t') 
  +  \frac{1}{T - t} \int_{0}^{t} dt' \,  \bar{I}(T) I(t') .
 \end{split}
\end{equation}
The ensemble average of Eq.~\eqref{ou-correlation-subtraction} is calculated to be
\begin{equation}
 \label{ou-correlation-subtraction-ea}
 \begin{split}
  \langle C'(t;T) \rangle
  & = e^{-t} - \langle \bar{I}^{2}(T) \rangle
  + \frac{1}{T - t}
  \bigg[ \int_{T - t}^{T} dt' \, \langle   \bar{I}(T) I(t') \rangle
  + \int_{0}^{t} dt' \, \langle   \bar{I}(T) I(t') \rangle \bigg] .
 \end{split}
\end{equation}
The ensemble averages appear in Eq.~\eqref{ou-correlation-subtraction-ea} are calculated as follows: 
\begin{equation}
\begin{split}
 \langle \bar{I}^{2}(T) \rangle
 & = \frac{1}{T^{2}} \int_{0}^{T} dt' \int_{0}^{T} dt'' \, \langle I(t') I(t'') \rangle 
  =  \frac{2}{T^{2}}  \int_{0}^{T} dt' \int_{0}^{t'} dt'' \, e^{- (t' - t'')} 
  =  \frac{2}{T^{2}}  (e^{-T} - 1 + T),
\end{split}
\end{equation}
\begin{equation}
\begin{split}
 \langle \bar{I}(T) I(t') \rangle
 & = \frac{1}{T}  \int_{0}^{T} dt'' \, \langle I(t') I(t'') \rangle
  = \frac{1}{T}  
 \left[ \int_{0}^{t'} dt'' \, e^{- t' + t''} 
 + \int_{t'}^{T} dt'' \, e^{- t'' + t'} 
 \right]
  = \frac{1}{T}  
 \left[ 2 - e^{-t'}
 - e^{- (T - t')}
 \right] ,
\end{split}
\end{equation}
\begin{equation}
 \begin{split}
  \int_{T - t}^{T} dt' \, \langle   \bar{I}(T) I(t') \rangle
  & = \int_{T - t}^{T} dt' \, \frac{1}{T}  
 \left[ 2 - e^{-t'}  - e^{- (T - t')} \right] 
   = \frac{1}{T} [2 t - 1 + e^{-T} + e^{-t} - e^{t - T}] ,
 \end{split}
\end{equation}
\begin{equation}
 \begin{split}
  \int_{0}^{t} dt' \, \langle   \bar{I}(T) I(t') \rangle
  & = \frac{1}{T} [2 t - 1 + e^{-T} + e^{-t} - e^{t - T}] .
 \end{split}
\end{equation}
Finally, we have the explicit expression for $\langle C'(t;T) \rangle$ as 
\begin{equation}
 \begin{split}
  \langle C'(t;T) \rangle
  & = e^{-t} - \frac{2 (T + t)}{(T - t) T^{2}}  (e^{-T} - 1 + T) 
  +  \frac{2}{T (T - t)} 
  [2 t - 1 + e^{-T} + e^{-t} - e^{t - T}].
 \end{split}
\end{equation}
Figure~\ref{ou_subtraction}(a) shows $\langle C'(t; T) \rangle$ with several different
$T$ values.
For comparison, $\langle C(t;T) \rangle = e^{-t}$ is also shown.
We find that $\langle C'(t;T) \rangle$ becomes zero at finite $t$.
This means that this correlation function apparently decays very rapidly.
Figure~\ref{ou_subtraction}(b) shows the same data as
Figure~\ref{ou_subtraction}(a), but in the semi-logarithmic scale.
We can observe that the shapes of $\langle C'(t;T) \rangle$ is not
largely affected by $T$ but $\langle C'(t;T) \rangle$ shifts to lower
values as $T$ decreases.

\begin{figure*}
\includegraphics[width=.8\linewidth, angle=0]{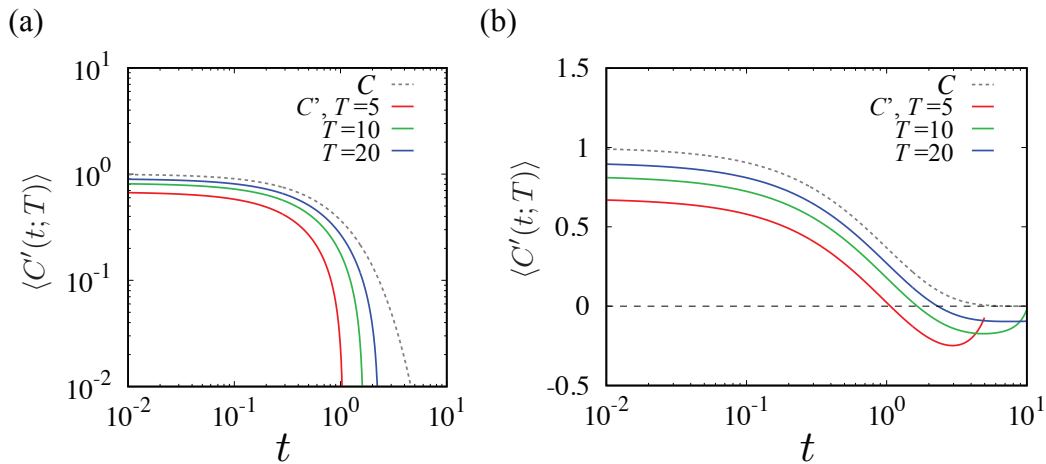}
 \caption{ (a) The time-averaged correlation functions of the Ornstein-Uhlenbeck process calculated by subtracting
 the time-average value in the observation time window. For comparison, the ensemble-averaged
 correlation function $C(t)=e^{-t}$ is shown (the broken curve).
 (b) The same data as  (a) but plotted in the semi-logarithmic scale.
}
 \label{ou_subtraction}
\end{figure*}
\end{widetext}

%

\end{document}